\documentclass[useAMS,usenatbib]{mn2e}
\usepackage[dvips]{graphicx}

\title[PSR J1829+2456: a relativistic binary pulsar]
      {PSR J1829+2456: a relativistic binary pulsar}
\author[D.~J.~Champion et al.]
{D.~J.~Champion$^{1}$,\thanks{Email: champion@jb.man.ac.uk} D.~R.~Lorimer$^{1}$, M.~A.~McLaughlin$^{1}$, J.~M.~Cordes$^{2}$,
\newauthor
Z.~Arzoumanian$^{3}$, J.~M.~Weisberg$^{4}$ \& J.~H.~Taylor$^{5}$\\
$^{1}$ University of Manchester, Jodrell Bank Observatory, Macclesfield, Cheshire, SK11 9DL\\ 
$^{2}$ Astronomy Department, Cornell University, Ithaca, NY 14853, USA\\ 
$^{3}$ USRA/LHEA, NASA Goddard Space Flight Center, Code 662, Greenbelt, MD 20771, USA\\
$^{4}$ Department of Physics and Astronomy, Carleton College, Northfield, MN 55057, USA\\ 
$^{5}$ Joseph Henry Laboratories and Physics Department, Princeton University, Princeton, NJ 08544, USA
}

\begin{document}

\date{\today}

\pagerange{\pageref{firstpage}--\pageref{lastpage}} \pubyear{2004}

\maketitle

\label{firstpage}

\begin{abstract}
We report the discovery of a new binary pulsar,  PSR~J1829+2456, found during a mid-latitude drift-scan survey with the Arecibo telescope. Our initial timing observations show the 41-ms pulsar to be in a 28-hr, slightly eccentric, binary orbit. The advance of periastron $\dot{\omega}=0.28\pm0.01$~deg yr$^{-1}$ is derived from our timing observations spanning 200 days. Assuming that the advance of periastron is purely relativistic and a reasonable range of neutron star masses for PSR~J1829+2456 we constrain the companion mass to be between 1.22~M$_\odot$ and 1.38~M$_\odot$, making it likely to be another neutron star. We also place a firm upper limit on the pulsar mass of 1.38~M$_\odot$.  The expected coalescence time due to gravitational-wave emission is long ($\sim$~60~Gyr) and this system will not significantly impact upon calculations of merger rates that are relevant to upcoming instruments such as LIGO.
\end{abstract}


\section{Introduction}

The first binary pulsar B1913+16 was discovered by Hulse \& Taylor (1975). This double neutron star (DNS) system with its 7.75-hr orbital period and large eccentricity ($e=0.6$) has since become a wonderful laboratory for testing general relativity in the strong-field regime \nocite{tw89} (Taylor \& Weisberg 1989). Perhaps most importantly, it has provided the first evidence for the existence of gravitational radiation \citep{tw82}. DNS binaries start life as binary systems with main-sequence stars of mass~$>$~6~M$_{\odot}$. Eventually the more massive of the two stars undergoes a supernova explosion, leaving a neutron star, sometimes seen as a pulsar. As the less-massive star evolves it increases in size until it overfills its Roche lobe. At this point, matter starts to accrete onto the pulsar causing it to spin up to periods as short as a few milliseconds \nocite{acrs82} (Alpar et al.~1982), a process known as recycling. In most cases the outer layers of the companion are blown away after accretion, exposing the core of the companion and producing a white dwarf-millisecond pulsar binary. For some recycled systems, however, the companion is massive enough to explode as a supernova. In most cases, this violent event will disrupt the binary system. The DNS binaries are those systems fortunate enough to survive. For further details see \cite{bv91}.

Our understanding of the DNS binary population is currently hampered by small-number statistics. Large-scale pulsar surveys are being carried out by a number of groups in order to improve this situation by increasing the sample of objects.  In this {\it Letter} we report on the results of a drift-scan survey using the Arecibo telescope. The survey was conducted at 430 MHz and covered mostly intermediate Galactic latitudes ($|b|<60^{\circ}$) to optimise the likelihood of finding millisecond and binary pulsars (see \nocite{cc97} e.g.~Cordes \& Chernoff 1997).  The known pulsars detected are described along with preliminary parameters for a new 41-ms pulsar J1829+2456 which is likely to be a DNS binary. We compare the properties of this new system with the known DNS and other relativistic binaries. Finally, we outline the future observational prospects for this system.

\section{Observations and Analysis}

In July and August of 1999 the Arecibo radio telescope was used to take the data presented here. The receiver was parked and observations were made as the sky drifted overhead. The newly commissioned 430-MHz receiver system in the Gregorian dome was used in combination with the Penn State Pulsar Machine (PSPM), a 128-channel analogue filterbank spectrometer which samples the incoming voltages from the telescope every 80~$\mu$s over a bandwidth of 7.68~MHz \citep{cad97b}. The Gregorian dome 430-MHz receiver has a gain of 11~K Jy$^{-1}$ and a system temperature of 45~K. At these latitudes a source passes through the beam in $\sim$~40~s. This is well matched to a $2^{19}$-sample time series collected with the PSPM (42~s). Our survey sparsely sampled a total of $\sim$~230~deg$^2$ within the right ascension range: $15 {\rm h} < \alpha < 20 {\rm h}$ for two declination ranges defined by $8^{\circ} < \delta < 12^{\circ}$ and $24^{\circ} < \delta < 28^{\circ}$. The instrumental setup was identical to that described by \nocite{lma+04} Lorimer et al.~(2004), in which the system sensitivity was estimated to be greater than or about 0.5~mJy for long period pulsars away from the Galactic plane and 3~mJy for millisecond pulsars. Full details of the search sensitivity will be given elsewhere.

The observations were analysed at the Jodrell Bank Observatory using a 182-processor Beowulf cluster (COBRA and an identical procedure to that described by Lorimer et \nocite{lma+04} al.~(2004). A suite of data analysis tools \citep{SIGPROC} was used in combination with scripts to keep each node continuously processing from the pool of data. Each drift observation was split into 42-s beams, with each beam overlapping 50\% of the previous beam. These beams were then dedispersed using trial dispersion measures (DMs) between 0 and 491.2~pc~cm$^{-3}$. The resulting time series were then Fourier transformed and searched for periodic signals. Incoherent summing of the first 2, 4, 8 and 16 harmonics was used to increase sensitivity to narrow pulse profiles. Any resulting candidates with a signal to noise ratio (S/N) greater than 8 were folded for visual inspection. It took $\sim$~13,000 CPU~hours to complete the analysis of the $\sim$~14,200 overlapping beams.

\section{Results and Follow-Up Observations}

In the $\sim$~230~deg$^{2}$ of sky covered by the survey, 4 known pulsars were detected; PSR~J1649+2533 (period, $P$~=~1.02~s; DM~=~34~pc~cm$^{-3}$) with S/N~$\sim$~80, PSR~J1652+2651 ($P$~=~0.92~s; DM~=~40.8~pc~cm$^{-3}$) with S/N~$\sim$~73, PSR~J1532+2745 ($P$~=~1.124~s; DM~=~14.7~pc~cm$^{-3}$) with S/N~$\sim$~38 and PSR~J1543+0929 ($P$~=~0.75~s; DM~=~35.2~pc~cm$^{-3}$) with S/N~$\sim$~85. The rate of pulsar detections in the data is consistent with the rate in previous drift--scan data taken using the line--feed in combination with the PSPM \nocite{mla+03} (McLaughlin et al.~2003).

\begin{figure}
\includegraphics[width=6cm, angle=270]{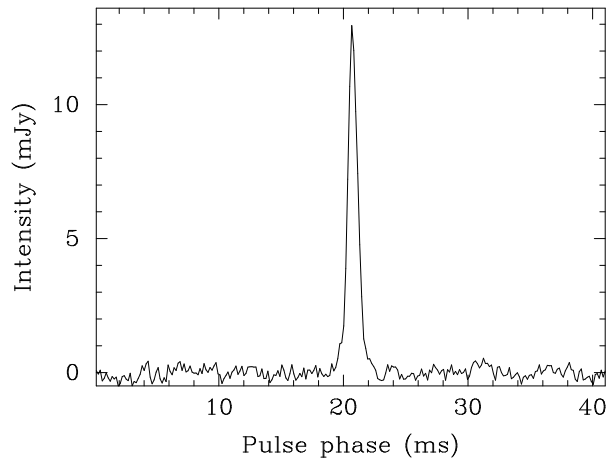}
\caption{Integrated pulse profile, formed from 4 hours of data, for PSR~J1829+2456 at 430-MHz.}
\label{fig:pulseprofile}
\end{figure}

One promising pulsar candidate was found with $P=41.02$~ms, DM $\sim$~13.9~pc~cm$^{-3}$ and S/N~$\sim$~10. The candidate was reobserved and confirmed as a new pulsar, J1829+2456, on June 3 2003 using the same observing system with a 10-min integration. Further observations showed that the pulsar scintillates strongly, with the flux density ranging from 0.07~mJy to 0.7~mJy. In Fig.~\ref{fig:pulseprofile} we present an integrated pulse profile. The confirmed approximate position put the pulsar in the line of sight of the Gould belt \citep{gmb+00,gre00a}, a dense region of atomic and molecular gas containing young stars and an ideal environment for producing strong scintillation. The DM from the search observation and the pulsar's Galactic coordinates ($l \simeq 53^{\circ}$; $b \simeq 15.6^{\circ}$) imply a distance of $\sim$~1.2~kpc \citep{cl02}. The true distance may in fact be much lower due to the unmodelled excess electron density in the Gould belt. The confirmation period was 40.99~ms, significantly shorter than the discovery period. This indicated that the pulsar was most likely a member of a binary system.

\begin{figure}
\includegraphics[width=6cm, angle=270]{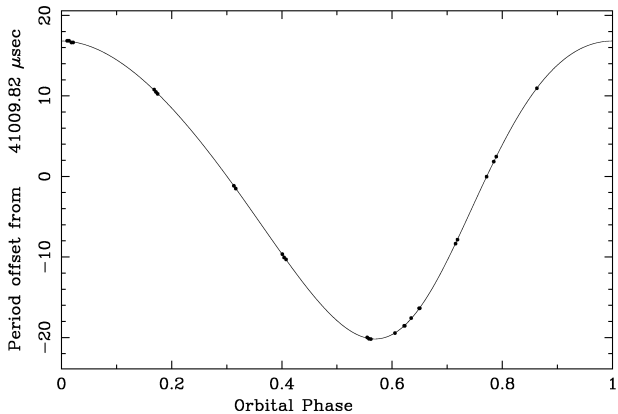}
\caption{The orbital fit (line) to the barycentric periods and times
(points) folded at the binary period of the system. The error bars are smaller than the points at this scale.}
\label{orbitfit}
\end{figure}

In order to determine the orbital parameters of PSR~J1829+2456, observations were made using the same observing system over the following 6 months. These data were dedispersed and searched for the period of the pulsar. Long observations when the pulsar appeared strongly were split and searched independently to provide more data points. The topocentric periods and observation times were then converted to barycentric periods and times using the TEMPO\footnote{http://pulsar.princeton.edu/tempo} software package. These barycentric periods and times were then fitted to possible binary orbits, resulting in the curve shown in Fig.~\ref{orbitfit}, which represents the pulse period evolution for an assumed orbital period $P_b \simeq 28$ hr, projected semi-major axis $x=a_{1}\sin~i \simeq 7.2$ lt-s and eccentricity $e \simeq 0.14$.

Using this preliminary ephemeris, the data were folded modulo the predicted period and cross correlated with a high S/N template profile to obtain an accurate time of arrival (TOA) for each observation. The TOAs were then analysed with TEMPO to produce a preliminary phase-connected solution (where every rotation of the pulsar is accounted for) over the time span of the observations. This resulted in the ephemeris given in Table~\ref{binpars}.

\section{Discussion}
\subsection{Mass determination}
\begin{figure}
\includegraphics[width=8cm, angle=0]{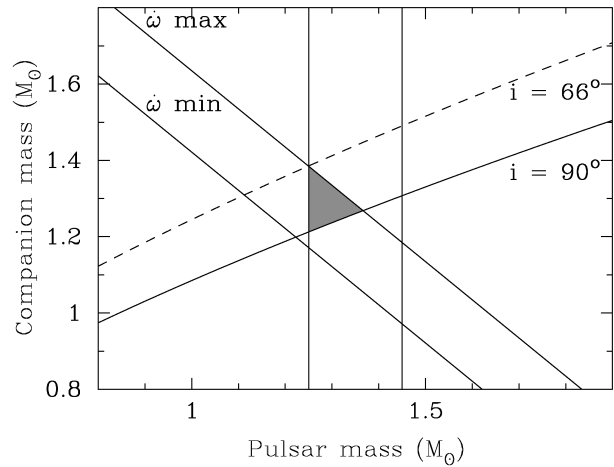}
\caption{The constraints on the masses of the pulsar and companion. The vertical lines are the mass limits for neutron stars described in Thorsett \& Chakrabarty (1999), and the low measured mass of PSR~J0737$-$3039B (Lyne et al. 2004). The lower companion mass limit is given by the mass function assuming a maximum inclination of $90^{\circ}$. The advance of periastron, $\dot{\omega}$, if assumed to be purely relativistic, provides the constraint for the maximum companion mass and maximum pulsar mass.}
\label{massdiag}
\end{figure}

The orbital parameters in Table~\ref{binpars} can be used to constrain the mass of the binary system. The Keplerian mass function relates the orbital period, $P_{b}$, and the projected semi-major axis, $x$, to the masses of the binary components. For this system we calculate a mass function 
\begin{equation}
f(m_1,m_2) = \frac{4\pi^{2} x^3}{P_b^2 T_{\odot}} = \frac{(m_{2}\sin i)^{3}}{(m_{1} + m_{2})^{2}} = 0.294\ \rm{M}_{\odot},
\label{cmass}
\end{equation}
where $T_{\odot} = GM_{\odot}c^{-3} = 4.925490947 \mu$s, $i$ is the inclination between the plane of the orbit and the line of sight, $x$ is in light-seconds, $P_b$ is in seconds and the pulsar and companion masses $m_1$ and $m_2$ are in Solar masses. Neutron stars are observed to have a narrow range of masses. Given the data in \cite{tc99}, and the low measured mass of PSR~J0737$-$3039B (Lyne et al. 2004), we expect the pulsar mass to lie between 1.25~M$_\odot$ and 1.47~M$_\odot$. For an orbit viewed edge-on ($i=90^{\circ}$), and a minimum neutron star mass for PSR~J1829+2456 of $m_{1}=1.25\ \rm{M}_\odot$, we calculate a minimum companion mass $m_2 \simeq$ 1.22 M$_\odot$. As shown in the following equation, the measurement of the advance of periastron, $\dot{\omega}$, if assumed to be purely relativistic, allows a measurement of the sum of the masses:
\begin{equation}
\dot\omega=3\left(\frac{2\pi}{P_{b}}\right)^{5/3}T_{\odot}^{2/3}(m_{1}+m_{2})^{2/3}(1-e^{2})^{-1}.
\label{omegadot}
\end{equation}
When the measured $\dot \omega$ is used in combination with the constraint that $\sin i<1$, we obtain an upper limit on the maximum pulsar mass~$m_1<1.38$ M$_{\odot}$. Similarly, assuming $m_1>1.25$ M$_{\odot}$, the maximum value of $\dot{\omega}$ implies $i>66^{\circ}$. These constraints are shown in a mass-mass diagram in Fig.~\ref{massdiag}. Although we cannot currently rule out a massive white dwarf or main-sequence star, given the similar mass functions, spin and orbital parameters of J1829+2456 to other DNS binaries (see Table~\ref{DNSsystems}) it seems most likely that the companion is another neutron star.

\subsection{A search for the companion}
The accuracy of the position of PSR~J1829+2456 inferred through radio timing is sufficient to allow for a search for an optical companion. No optical companion is present in the uncalibrated plates of the Digitised Sky Survey. Preparations are underway for a deep optical search for the companion star.

Given the possibility that the companion of J1829+2456 is a neutron star, a search was made of the available radio data for a periodic signal which could be coming from the companion if it were active as a radio pulsar. This search was given greater significance following the recent detection of 2.8-s pulsations from the companion of PSR~J0737$-$3039A (Lyne et \nocite{lbk+04} al.~2004). Initial searches of the data (at the DM of PSR~J1829+2456) looking for other periodic signals showed that, if the companion was an active radio pulsar, it was too weak to be easily detectable. 

To carry out a more sensitive search for the companion, we made use of the fact that the orbital parameters of the system are known and that the pulsations from the companion star would be Doppler shifted in the opposite sense to PSR~J1829+2456 due to its orbital motion about their common centre of mass. To correct for this effect, we applied the first-order Doppler formula to calculate the effective sampling interval in the rest-frame of the companion:
\begin{equation}
  t_{\rm companion} = t_{\rm samp} (1 + v(t)/c),
\end{equation}
where $t_{\rm samp}$ is the sampling interval at the observatory and $v(t)$ is the effective line-of-sight radial velocity of the companion as a function of time, $t$. We used TEMPO to calculate $v(t)$ to account for the effects of orbital motion in the binary system (using the orbital parameters given in Table~\ref{binpars}, assuming a pulsar mass of 1.35~M$_\odot$ and changing only the longitude of periastron by $180^{\circ}$) as well as contributions from the Earth's rotation and motion about the Sun. All available data were dedispersed and resampled in this way before being passed through the standard search analysis. No significant candidates were found down to a S/N of 5. This corresponds to a 430-MHz flux limit of $\sim$~0.1~mJy.

While the lack of detection of the companion suggests that it is too weak to be seen as a radio pulsar, or unfavourably beamed, it should be noted that the companion to PSR~J0737$-$3039A is only clearly detectable over certain phases of the orbit (Lyne et al.~2004). If any companion to PSR~J1829+2456 was behaving in a similar fashion it may not be visible in the current data set. Further, more sensitive, searches will be carried out as more data are taken.

\subsection{Expected relativistic parameters}
Using the constrained range of masses for the pulsar and companion the expected orbital period derivative and period for geodetic precession due to misalignment of the spin axis of the pulsar with the orbital angular momentum vector \citep{bo75b} can be determined. Assuming a combined mass of 2.53~M$_\odot$, the expected values for $\dot P_{b}$ and the expected timescale for geodetic precession are given in Table~\ref{binpars}. The expected $\gamma$ should be measurable after 18 months of timing data. The expected $\dot P_{b}$ is too small to be measurable without a long-term (5--10 yr) timing campaign. The expected timescale for geodetic precession is $\sim$~4200~yrs, too long to be measurable through pulse profile or timing variations. 

\begin{table}
\caption{Measured, derived, and expected parameters for PSR~J1829+2456 from 29 TOAs from MJDs 52786 to 52986.}
\begin{center}
\begin{tabular}{l l}
\hline
\multicolumn{2}{c}{Measured Parameters}                                         \\
\hline
Right ascension (J2000) (h:m:s)                                & 18:29:34.6 (1) \\
Declination (J2000) ($^{\circ}$:':'')                          & 24:56:19 (2)   \\
Period (ms)                                                    & 41.00982358 (1)\\
Epoch of period (MJD)                                          & 52887          \\
Projected semi-major axis (s)                                  & 7.2360 (1)     \\
Dispersion measure (pc cm$^{-3}$)                              & 13.9 (5)       \\
Binary period (days)                                           & 1.176028 (1)   \\
Eccentricity                                                   & 0.13914 (4)    \\
Longitude of periastron (deg)                                  & 229.94 (1)     \\
Epoch of periastron (MJD)                                      & 52848.57977 (3)\\
430-MHz flux density (mJy)                                     & 0.3 (1)        \\
Pulse width at 50\% of peak $w_{50}$ (ms)                      & 1.07 (16)      \\
Pulse width at 10\% of peak $w_{10}$ (ms)                      & 2.07 (16)      \\
Rate of advance of periastron (deg yr$^{-1}$)                  & 0.28 (1)       \\
Mass function (M$_\odot$)                                      & 0.29413 (1)    \\
RMS residual to fit ($\mu$s)                                   & 19             \\
\hline
\multicolumn{2}{c}{Derived Parameters}                                          \\
\hline
Galactic longitude (J2000) (deg)                               & 53.343 (1)     \\
Galactic latitude (J2000) (deg)                                & 15.612 (1)     \\
Distance (kpc)$^{a}$                                           & $\sim$ 1.2     \\
Mean orbital speed (km s$^{-1}$)                               & 134            \\
Total system mass (M$_\odot$)                                  & 2.5 (2)        \\
Minimum companion mass (M$_\odot$)                             & 1.22           \\
Maximum pulsar mass (M$_\odot$)                                & 1.38           \\
\hline
\multicolumn{2}{c}{Expected Relativistic Parameters}                            \\
\hline
Orbital period derivative ($\times10^{-12}$)                   & $-$0.02        \\
Coalescence time (Gyr)                                         & 60             \\
Geodetic precession rate (deg yr$^{-1}$)                       & 0.075          \\
Relativistic time-dilation\\ and gravitational redshift $\gamma$ (ms)   & 1.3   \\
\hline
\end{tabular}
\end{center}
a: Distance inferred from DM \citep{cl02}.\\
The numbers in parentheses are the 1--$\sigma$ uncertainties in the least significant digit quoted.
\label{binpars}
\end{table}

\subsection{Comparison with other relativistic binary systems}
If our future measurements show PSR~J1829+2456 to be a double neutron star system it will be the seventh such system to be discovered. Table~\ref{DNSsystems} contains a list of the most relativistic binary pulsar systems and their orbital parameters. We also list the coalescence time $\tau_{\rm GW}$ due to the emission of gravitational radiation of each system. Following Lorimer (2001)\footnote{We reproduce the formula here to correct a typographic error in the original citation.}, we can approximate the detailed calculations of $\tau_{\rm GW}$ by Peters \nocite{pet64} (1964) via
\begin{equation}
\label{coaltime}
\tau_{\rm GW}     \simeq  10^7 \, {\rm yr} \,
                  \left(\frac{P_{b}}{\rm hr}\right)^{8/3}
                  \left(\frac{\mu}{{\rm M}_{\odot}}\right)^{-1}
		  \left(\frac{m_1+m_2}{{\rm M}_{\odot}}\right)^{-2/3}
		  (1-e^{2})^{7/2},
\end{equation}
where the reduced mass $\mu = m_1 m_2 / (m_1+m_2)$. For any reasonable
range of $m_1$ and $m_2$ discussed above, we find $\tau_{\rm GW} \sim$~60~Gyr.

The systems listed in Table~\ref{DNSsystems} span a wide range of orbital parameters but can be broadly split into systems that will and will not coalesce within a Hubble time, with PSR~J1829+2456 lying toward the edge of the non-coalescing group. The relatively long coalescence time for J1829+2456 means that it will not affect the DNS merger rate calculations (e.g. Kalogera et al. 2004) for gravitational wave detectors such as LIGO.

\begin{table*}
\caption{The orbital parameters of various eccentric binary systems.}
\begin{center}
\begin{tabular}{l r@{.}l r@{.}l r@{.}l r@{.}l r@{.}l r@{.}l c r@{.}l c c}
\hline
PSR &\multicolumn{2}{c}{$P$} &\multicolumn{2}{c}{$P_{b}$} &\multicolumn{2}{c}{$a_{1}\sin~i$} &\multicolumn{2}{c}{$e$} &\multicolumn{2}{c}{$\dot \omega$} &\multicolumn{2}{c}{$\dot P _{b}$} & $f(m)$ &\multicolumn{2}{c}{$m_{1} + m_{2}$} &$\tau_{\rm GW}^{\dag}$ & References\\
&\multicolumn{2}{c}{(ms)}  &\multicolumn{2}{c}{(days)} &\multicolumn{2}{c}{(lt-s)} &\multicolumn{2}{c}{} &\multicolumn{2}{c}{(deg yr$^{-1}$)} &\multicolumn{2}{c}{($\times 10^{-12}$)} & \multicolumn{2}{c}{(M$_\odot$)} &(M$_\odot$) &(Gyr) & \\
\hline
&\multicolumn{17}{c}{Double neutron star binaries}\\
\hline
B1913+16     &  59&03 &  0&323 &  2&34 & 0&617 &  4&227                        & $-$2&428                 & 0.13 & 2&83                   & 0.31  & 1  \\
B1534+12     &  37&90 &  0&421 &  3&73 & 0&274 &  1&756                        & $-$0&138                 & 0.31 & 2&75                   & 2.69  & 2  \\
B2127+11C    &  30&53 &  0&335 &  2&52 & 0&681 &  4&457                        & $-$3&937                 & 0.15 & 2&71                   & 0.22  & 3  \\
J1518+4904   &  40&93 &  8&634 & 20&04 & 0&249 &  0&011                        &\multicolumn{2}{c}{--}    & 0.12 & 2&62                   & 9600  & 4  \\
J1811$-$1736 & 104&18 & 18&779 & 34&78 & 0&828 &  0&009                        &\multicolumn{2}{c}{$<$30} & 0.13 & 2&6                    & 1700  & 5  \\
J0737$-$3039A&  22&70 &  0&102 &  1&42 & 0&088 & 16&88                         & $-$1&24$^{\ast}$         & 0.29 & 2&58                   & 0.087 & 6  \\
\hline
&\multicolumn{17}{c}{White dwarf binaries}\\
\hline
B2303+46     &1066&37 & 12&34  & 32&69 & 0&66  &  0&010                        &\multicolumn{2}{c}{--}    & 0.25 & 2&53                   & 4500  & 7  \\
J1141$-$6545 & 393&90 &  0&20  &  1&86 & 0&17  &  5&33                         &\multicolumn{2}{c}{$<$50} & 0.18 & 2&30                   & 0.59  & 8  \\
\hline
&\multicolumn{17}{c}{Unknown companion}\\
\hline
B1820$-$11   & 279&83 &357&76  &200&67 & 0&79  &\multicolumn{2}{c}{$<10^{-4}$} &\multicolumn{2}{c}{--}    & 0.07 & \multicolumn{2}{l}{--} & --    & 9  \\
J1829+2456   &  41&00 &  1&17  &  7&24 & 0&14  &  0&28                         & $-$0&02$^{\ast}$         & 0.29 & 2&53                   & 60    & -- \\
\hline
\end{tabular}
\begin{tabular}{l l}
1: \cite{ht75a, wt03a, tw89} &6: \cite{bdp+03}                                \\
2: \cite{wol90z, sttw02}     &7: \cite{tamt93, arz95}                         \\
3: \cite{agk+90, and92}      &8: \cite{klm+00a}                               \\
\cite{pakw91, dk96}          &9: \cite{lm89, tamt93}                          \\
4: \cite{nst96, hlk+03}      &$\ast$: Predicted value.                        \\
5: \cite{lcm+00}             &$\dag$: Calculated using formula \ref{coaltime}.\\
\end{tabular}
\end{center}
\label{DNSsystems}
\end{table*}

\section{Future Study}
In an Arecibo drift-scan survey of 230 deg$^2$ we have discovered a new relativistic binary system, PSR~J1829+2456. Our measurements so far suggest that the companion is most likely another neutron star.The long predicted gravitational-wave coalescence time means that this system has no effect on the calculations of merger rates required for instruments such as LIGO. A dedicated Arecibo timing campaign will allow more accurate measurements of the position of this system, the spin-down rate of the pulsar and the relativistic advance of periastron. Continued timing will provide additional valuable information: the system's proper motion, the relativistic time-dilation and gravitational redshift parameter $\gamma$, and possibly a measurement of Shapiro delay. An optical search for the companion is planned to confirm whether it is a massive white dwarf or, as seems most likely, another neutron star. Studies of scintillation effects are also planned which should allow the inclination angle of the orbit to be determined.

\section*{Acknowledgements}
The Arecibo observatory, a facility of the National Astronomy and Ionosphere Center, is operated by Cornell University in a co-operative agreement with the National Science Foundation (NSF).  We thank Alex Wolszczan for making the PSPM freely available for use at Arecibo. Without this superb instrument, the results presented here would not have been possible. DJC is funded by the Particle Physics and Astrophysics Research Council. DRL is a University Research Fellow funded by the Royal Society. ZA was supported by NASA grant NRA-99-01-LTSA-070.

\bibliographystyle{mn2e}
\bibliography{journals.bib,psrrefs.bib,modrefs.bib,myrefs.bib}

\begin{thebibliography}{}

\bibitem[\protect\citeauthoryear{Alpar, Cheng, Ruderman \& Shaham}{Alpar
  et~al.}{1982}]{acrs82}
Alpar M.~A.,  Cheng A.~F.,  Ruderman M.~A.,    Shaham J.,  1982, Nature, 300,
  728

\bibitem[\protect\citeauthoryear{Anderson}{Anderson}{1992}]{and92}
Anderson S.~B.,  1992, PhD thesis, California Institute of Technology

\bibitem[\protect\citeauthoryear{Anderson, Gorham, andT. A.~Prince \&
  Wolszczan}{Anderson et~al.}{1990}]{agk+90}
Anderson S.~B.,  Gorham P.~W.,  andT. A.~Prince S. R.~K.,    Wolszczan A.,
  1990, Nature, 346, 42

\bibitem[\protect\citeauthoryear{Arzoumanian}{Arzoumanian}{1995}]{arz95}
Arzoumanian Z.,  1995, PhD thesis, Princeton University

\bibitem[\protect\citeauthoryear{{Barker} \& {O'Connell}}{{Barker} \&
  {O'Connell}}{1975}]{bo75b}
{Barker} B.~M.,  {O'Connell} R.~F.,  1975, Phys. Rev. D, 12, 329

\bibitem[\protect\citeauthoryear{Bhattacharya \& {van den Heuvel}}{Bhattacharya
  \& {van den Heuvel}}{1991}]{bv91}
Bhattacharya D.,  {van den Heuvel} E. P.~J.,  1991, Phys. Rep., 203, 1

\bibitem[\protect\citeauthoryear{{Burgay}, {D'Amico}, Possenti, Manchester,
  Lyne, Joshi, {McLaughlin}, Kramer, Sarkissian, Camilo, Kalogera, Kim \&
  Lorimer}{{Burgay} et~al.}{2003}]{bdp+03}
{Burgay} M.,  {D'Amico} N.,  Possenti A.,  Manchester R.~N.,  Lyne A.~G.,
  Joshi B.~C.,  {McLaughlin} M.,  Kramer M.,  Sarkissian J.~M.,  Camilo F.,
  Kalogera V.,  Kim C.,    Lorimer D.~R.,  2003, Nature, 426, 531

\bibitem[\protect\citeauthoryear{{Cadwell}}{{Cadwell}}{1997}]{cad97b}
{Cadwell} B.~J.,  1997, PhD thesis, Pennsylvania State University

\bibitem[\protect\citeauthoryear{Cordes \& Chernoff}{Cordes \&
  Chernoff}{1997}]{cc97}
Cordes J.~M.,  Chernoff D.~F.,  1997, ApJ, 482, 971

\bibitem[\protect\citeauthoryear{{Cordes} \& {Lazio}}{{Cordes} \&
  {Lazio}}{2002}]{cl02}
{Cordes} J.~M.,  {Lazio} T.~J.~W.,  2002, ApJ

\bibitem[\protect\citeauthoryear{Deich \& Kulkarni}{Deich \&
  Kulkarni}{1996}]{dk96}
Deich W. T.~S.,  Kulkarni S.~R.,  1996, in van Paradijs J.,  van~del Heuvel E.
  P.~J.,   Kuulkers E.,  eds, {C}ompact {S}tars in {B}inaries: {IAU} Symposium
  165 The masses of the neutron stars in {M15C}.
Kluwer, Dordrecht, pp 279--285

\bibitem[\protect\citeauthoryear{{Gehrels}, {Macomb}, {Bertsch}, {Thompson} \&
  {Hartman}}{{Gehrels} et~al.}{2000}]{gmb+00}
{Gehrels} N.,  {Macomb} D.~J.,  {Bertsch} D.~L.,  {Thompson} D.~J.,
  {Hartman} R.~C.,  2000, Nature, 404, 363

\bibitem[\protect\citeauthoryear{Grenier}{Grenier}{2000}]{gre00a}
Grenier I.~A.,  2000, Astron. Astrophys, 364, L93

\bibitem[\protect\citeauthoryear{Hobbs, Lyne, Kramer, Martin \& Jordan}{Hobbs
  et~al.}{2003}]{hlk+03}
Hobbs G.,  Lyne A.~G.,  Kramer M.,  Martin C.~E.,    Jordan C.,  2003, MNRAS

\bibitem[\protect\citeauthoryear{Hulse \& Taylor}{Hulse \&
  Taylor}{1975}]{ht75a}
Hulse R.~A.,  Taylor J.~H.,  1975, ApJ, 195, L51

\bibitem[\protect\citeauthoryear{{Kaspi}, {Lyne}, {Manchester}, {Crawford},
  {Camilo}, {Bell}, {D'Amico}, {Stairs}, {McKay}, {Morris} \&
  {Possenti}}{{Kaspi} et~al.}{2000}]{klm+00a}
{Kaspi} V.~M.,  {Lyne} A.~G.,  {Manchester} R.~N.,  {Crawford} F.,  {Camilo}
  F.,  {Bell} J.~F.,  {D'Amico} N.,  {Stairs} I.~H.,  {McKay} N. P.~F.,
  {Morris} D.~J.,    {Possenti} A.,  2000, ApJ, 543, 321

\bibitem[\protect\citeauthoryear{Lorimer}{Lorimer}{2001}]{SIGPROC}
Lorimer D.~R.,  2001, (Arecibo Tech. Operations Memo. 01-01) (Arecibo: Natl.
  Astron. Ionos. Cent.)

\bibitem[\protect\citeauthoryear{Lorimer, McLaughlin, Arzoumanian, Xilouris,
  Cordes, Lommen, Fruchter, Chandler \& Backer}{Lorimer et~al.}{2004}]{lma+04}
Lorimer D.~R.,  McLaughlin M.~A.,  Arzoumanian Z.~A.,  Xilouris K.,  Cordes
  J.~M.,  Lommen A.~N.,  Fruchter A.~S.,  Chandler A.~M.,    Backer D.~C.,
  2004, MNRAS, 347, L21

\bibitem[\protect\citeauthoryear{Lyne, {Burgay}, Kramer, Possenti, Manchester,
  Camilo, {McLaughlin}, Lorimer, Joshi, Reynolds \& Freire}{Lyne
  et~al.}{2004}]{lbk+04}
Lyne A.~G.,  {Burgay} M.,  Kramer M.,  Possenti A.,  Manchester R.~N.,  Camilo
  F.,  {McLaughlin} M.,  Lorimer D.~R.,  Joshi B.~C.,  Reynolds J.~E.,
  Freire P. C.~C.,  2004, Science, 303, 1153

\bibitem[\protect\citeauthoryear{Lyne, Camilo, Manchester, Bell, Kaspi,
  D'Amico, McKay, Crawford, Morris, Sheppard \& Stairs}{Lyne
  et~al.}{2000}]{lcm+00}
Lyne A.~G.,  Camilo F.,  Manchester R.~N.,  Bell J.~F.,  Kaspi V.~M.,  D'Amico
  N.,  McKay N. P.~F.,  Crawford F.,  Morris D.~J.,  Sheppard D.~C.,    Stairs
  I.~H.,  2000, MNRAS, 312, 698

\bibitem[\protect\citeauthoryear{Lyne \& McKenna}{Lyne \& McKenna}{1989}]{lm89}
Lyne A.~G.,  McKenna J.,  1989, Nature, 340, 367

\bibitem[\protect\citeauthoryear{McLaughlin, Lorimer, Arzoumanian, Backer,
  Cordes, Fruchter, Lommen \& Xilouris}{McLaughlin et~al.}{2003}]{mla+03}
McLaughlin M.~A.,  Lorimer D.~R.,  Arzoumanian Z.~A.,  Backer D.~C.,  Cordes
  J.~M.,  Fruchter A.~S.,  Lommen A.~N.,    Xilouris K.,  2003, in M.~Bailes
  D.~Nice .~S.~T.,  ed., Radio Pulsars (ASP Conf.~Ser. New pulsars from arecibo
  drift searches.
PASP, San Francisco, pp 129--132

\bibitem[\protect\citeauthoryear{Nice, Sayer \& Taylor}{Nice
  et~al.}{1996}]{nst96}
Nice D.~J.,  Sayer R.~W.,    Taylor J.~H.,  1996, ApJ, 466, L87

\bibitem[\protect\citeauthoryear{Peters}{Peters}{1964}]{pet64}
Peters P.~C.,  1964, Phys. Rev., 136, 1224

\bibitem[\protect\citeauthoryear{Prince, Anderson, Kulkarni \&
  Wolszczan}{Prince et~al.}{1991}]{pakw91}
Prince T.~A.,  Anderson S.~B.,  Kulkarni S.~R.,    Wolszczan W.,  1991, ApJ,
  374, L41

\bibitem[\protect\citeauthoryear{Stairs, Thorsett, Taylor \& Wolszczan}{Stairs
  et~al.}{2002}]{sttw02}
Stairs I.~H.,  Thorsett S.~E.,  Taylor J.~H.,    Wolszczan A.,  2002, ApJ, 581,
  501

\bibitem[\protect\citeauthoryear{Taylor \& Weisberg}{Taylor \&
  Weisberg}{1982}]{tw82}
Taylor J.~H.,  Weisberg J.~M.,  1982, ApJ, 253, 908

\bibitem[\protect\citeauthoryear{Taylor \& Weisberg}{Taylor \&
  Weisberg}{1989}]{tw89}
Taylor J.~H.,  Weisberg J.~M.,  1989, ApJ, 345, 434

\bibitem[\protect\citeauthoryear{Thorsett, Arzoumanian, McKinnon \&
  Taylor}{Thorsett et~al.}{1993}]{tamt93}
Thorsett S.~E.,  Arzoumanian Z.,  McKinnon M.~M.,    Taylor J.~H.,  1993, ApJ,
  405, L29

\bibitem[\protect\citeauthoryear{Thorsett \& Chakrabarty}{Thorsett \&
  Chakrabarty}{1999}]{tc99}
Thorsett S.~E.,  Chakrabarty D.,  1999, ApJ, 512, 288

\bibitem[\protect\citeauthoryear{{Weisberg} \& {Taylor}}{{Weisberg} \&
  {Taylor}}{2003}]{wt03a}
{Weisberg} J.~M.,  {Taylor} J.~H.,  2003, in ASP Conf. Ser. 302: Radio Pulsars
  {The Relativistic Binary Pulsar B1913+16}.
p.~93

\bibitem[\protect\citeauthoryear{{Wolszczan}}{{Wolszczan}}{1990}]{wol90z}
{Wolszczan} A.,  1990, in International Astronomical Union Circular {PSR
  1257+12 and PSR 1534+12}.
p.~1

\end{thebibliography}

\label{lastpage}

\end{document}